\documentclass{acm_proc_article-sp}

\usepackage[utf8]{inputenc}

\usepackage{longtable}

\usepackage{pdflscape}

\usepackage{graphicx}
\graphicspath{ {../graphics/} }
\usepackage{float}

\begin{document}

\title{Exposing the Hidden Web: An Analysis of Third-Party HTTP Requests on One Million Websites
}
\subtitle{\emph{International Journal of Communication, October 2015.}}
%
%
%
%
%

\numberofauthors{1} 
%
\author{
%
%
\alignauthor
Timothy Libert\titlenote{\\ public@timlibert.me \\ https://timlibert.me \\ \\ Annenberg School for Communication\\University of Pennsylvania \\ 3620 Walnut Street \\ Philadelphia, PA, USA}\\
}

\maketitle

\begin{abstract}
This article provides a quantitative analysis of privacy compromising mechanisms on one million popular websites. Findings indicate that nearly nine in ten websites leak user data to parties of which the user is likely unaware of; over six in ten websites spawn third-party cookies; and over eight in ten websites load Javascript code from external parties onto users’ computers. Sites which leak user data contact an average of nine external domains, indicating users may be tracked by multiple entities in tandem. By tracing the unintended disclosure of personal browsing histories on the web, it is revealed that a handful of American companies receive the vast bulk of user data. Finally, roughly one in five websites are potentially vulnerable to known NSA spying techniques at the time of analysis.
\end{abstract}


\terms{Behavioral Tracking, Hidden Web, Privacy, Advertising, Internet Policy, Do Not Track}



%


\section{Introduction}

		\emph{If, as it seemed, the new technology was on a collision course with the values of personal privacy and human dignity, could the collision be averted? Could a system be devised to identify and permit beneficial uses of the new technology and yet, at the same time, preclude those uses that most men would deem intolerable?}
	
	\begin{flushright}
		- Alan Westin, 1967
	\end{flushright}

	As the past several decades have shown, the beneficial aspects of information technologies have proven truly remarkable and fundamentally transformative, yet they have also ushered in an age of nearly limitless surveillance.  It is this use of information technology for surveillance purposes which Alan Westin described as potentially ``intolerable" nearly 50 years ago \cite{westin-1967-privacy}.  Indeed, every new device, app, and social network is now assumed to come with hidden privacy risks.  Where the economy of the past century relied on cash for most transactions; personal behavioral data has become a currency of its own in the current century.  Learning what an individual does with her or his time, be it browse specific websites or purchase particular products, has become a means to more efficiently target products and services to those who fit certain behavioral profiles.
	
	Efforts to collect personal behavioral data has extended to governments as well. News reports based on internal documents of the United States National Security Agency (NSA) leaked by former contractor Edward Snowden have revealed that the U.S. government has constructed a surveillance system of such complexity and scope that it fulfills the dire warnings made by Senator Frank Church, who suggested in 1975 that the NSA's technological capabilities carried with them ``the capacity...to make tyranny total in America", and that a fully realized technological surveillance state constituted ``abyss from which there is no return" \cite{bamford-2011-enemies}.  Thus, looking back on the worries of both Westin and Church within the context of the present day, we may see that their warnings, however extreme, had significant merit.  It is now clear that information technologies carry with them externalities which are socially intolerable, and that the realities of state surveillance often exceed the worst predictions of yesteryear.  However, despite the warnings of Westin and Church proving prescient, there are many opportunities to shift the balance of power in favor of privacy and user choice.

	While Westin and Church both dealt with many hypothetical scenarios and little knowledge of future technologies, today the general principals of computing, data storage, and networking are well known.  Those with technical knowledge of computing systems may use that knowledge to analyze the surveillance systems which affect the lives of billions of Internet users.  As with discussions of other social ills such as poverty and preventable disease, debates on issues of surveillance are served when backed by sound quantitative measures.  Thus, in order to escape Church's ``abyss", public policy advocates must identify and enumerate specific forms of surveillance, the scope of this surveillance, and the commercial and government practitioners of it.
	
	The purpose of this article is to provide a quantitative analysis of privacy compromising mechanisms on the top one million websites as determined by the Alexa company.  It shall be demonstrated that nearly nine in ten websites leak user data to parties of which the user is likely unaware of; over six in ten websites spawn third-party cookies; and over eight in ten websites load Javascript code.  Sites which leak user data contact an average of nine external domains.  Most importantly, by tracing the flows of personal browsing histories on the web, it is possible to discover the corporations which profit from tracking user.  While there are a number of companies tracking users online, the overall landscape is highly consolidated, with the top corporation, Google, tracking users on nearly eight of ten sites in the Alexa top one million.  Finally, by consulting internal NSA documents leaked by Edward Snowden, it has been determined that roughly one in five websites are potentially vulnerable to known NSA spying techniques at the time of analysis.

	It should be noted that this is not the first quantitative study of privacy online.  However, the vast majority of quantitative research in the area of web tracking has appeared in the computer science literature with scholars such as Krishnamurty and Wills conducting foundational studies as far back as 2006.  These studies provide essential insights into the nature and scope of online tracking, but require a high level of technical expertise to understand.  This has had the unfortunate side-effect of re-enforcing barriers between disciplines and reducing the potential for cross-disciplinary work. This is despite the fact the Communication discipline has traditionally played a strong role in areas of telecommunications and media policy.  For these reasons, this article aims to bridge a disciplinary gap.

	This article begins with a technical primer on how web tracking works, explores previous research in the area, and then explicates several research questions.  This is followed by a methodology section which details a new open-source software platform, named webXray, which has been developed to conduct this research.  Finally, findings are presented along with suggestions for how policy makers may shift the balance away from unchecked surveillance towards respect for user preferences.

\section{Background: Third-Party HTTP Requests and User Tracking}

	In order to fully appreciate how companies track users on the web, a technical aside is in order.  When a user's web browser makes a request to load a web page it receives an HTML file from the server associated with a given domain (such as ``http://example.com" where ``example.com" is the domain).  Any request to download content, such as a web page, uses the hypertext transfer protocol (HTTP) in order to connect the user to the server hosting the content they desire \cite{fielding-1999-http}.  There are only two parties involved in the transaction: the user and the company who owns server.  Due to the fact that only one server is being connected to, and the address of that server is visible in the browser address bar, these are often called ``first-party" requests.

	    The beauty of HTML, and the web itself, is that HTML pages can include additional pieces of content, called ``elements", from a limitless array of additional parties.  Such elements may include pictures, videos, and Javascript code.  However, when new content is included from outside domains, the number of parties involved goes from two to three: the user, the site she or he is visiting, and the entity providing the additional content.  For this reason, an image or other piece of content from an external server may be called a ``third-party element" (sometimes abbreviated to ``3PE"), and the process of downloading such an element is the result of a ``third-party request". 
	    
	Every time an HTTP request is made, information about the user is transmitted to the server hosting the content.  This data includes the IP address of the computer making the request, the date and time the request was made, as well as the type of computer and web browser employed by the user, which is known as the ``user-agent" field.  In addition, the address of the page which initiated the request, known as the ``referer" [sic], is included.  This information is found in the raw data of the HTTP request itself, an example of which is shown below:

	\begin{verbatim}
		IP: 8.67.53.09
		DATE: [10/May/2014:19:54:25 +0000] 
		REQUEST: "GET /tracking_pixel.png HTTP/1.1"
		REFERER: "http://example.com/private_matters.html" 
		USER-AGENT: "Mac OS X 10_8_5...AppleWebKit/537.71"
	\end{verbatim}

	From the above information, the server receiving the request for the file ``tracking\_pixel.png" can determine that it was made from a user with the IP address 8.67.53.09, using the Safari browser on a Macintosh computer, who is currently viewing a webpage with the address ``http://example.com/ private\_matters.html".  If the server has many such records, patterns of behavior may be attributed to the same combination of IP and user-agent information.  This is the most basic form of tracking and is common to all HTTP requests made on the web.  
	
	Before companies engaged in web tracking may harvest HTTP request data they must first get their elements included in the source code of a page.  In some cases advertisers pay to have their elements on a page, and this is the dominant business model for online publishers \cite{turow-2012-dailyyou}.  However, the majority of third-party HTTP requests are made to corporations such as Google, Facebook, and Twitter which provide free services such as traffic analytics and social media tools.  While web masters do not pay for such tools, the hidden cost of privacy loss is incurred by unwitting users.

	Despite the fact that this situation is already rather murky from the user's perspective, it actually gets worse.  Some third-party elements are also able to store information on a user's computer using small files called \emph{cookies} \cite{kristol-1997-http_state_mangement_rfc}.  Cookies may be used as a unique identifier which is affixed to the browser of a given user.  This is analogous to the tracking bracelets ornithologists place on migratory birds, even if users visit unrelated sites, tracking cookies have rendered them tagged and traceable.  A user may delete or block cookies, potentially restoring some privacy.  However, other third-party requests may download sophisticated Javascript code which can execute commands on users' computers, facilitating advanced forms of tracking known as ``browser fingerprinting" (see Eckersley, 2010 for an early study on the topic).  By detecting the unique characteristics of a given computer, browser fingerprinting is analogous its offline form; the specific list of fonts installed on a computer may function like a loop or a whorl on a finger - given combinations are often unique to an individual.  Also like real fingerprints, it is often difficult to hide or change a browser's fingerprint.  The attention of corporations is rapidly turning to fingerprinting techniques for this reason and cookies are quickly falling out of favor.

\section{Previous Research}

	Previous research in web privacy has generally focused on three main areas: attitudes of users towards the idea of being tracked, the mechanisms which facilitate tracking, and measurement of how much tracking is happening.  The majority of research regarding attitudes has taken place within the fields of communication and law, whereas mechanisms and measurement research has primarily been the domain of computer science.

	\subsection{Attitudes}
	Opinion research regarding online privacy has been ongoing since the 1990s.  Yet, as much as the web has changed since that decade, attitudes towards privacy have remained fairly constant.  In 1999 ``respondents registered a high level of concern about privacy in general and on the Internet" \cite{ackerman-1999-privacy}.  Likewise, a 2003 study found that ``a clear majority of Americans express worry about their personal information on the web" \cite{turow-2003-americans}.  A 2009 survey found that ``69\% of American adults fe[lt] there should be a law that gives people the right to know everything that a website knows about them" \cite{turow-2009-americans}.  Similar work in 2012 showed that 60\% of respondents would like the proposed ``Do Not Track" (DNT) standard to prevent websites from collecting personal information (DNT is further explained in the discussion section) \cite{hoofnagle-2012-privacy}.  In a 2014 study, ``Public Perceptions of Privacy and Security in
the Post-Snowden Era", the Pew Research Center discovered that 70\% of respondents felt records of the websites they had visited constituted ``very" or ``somewhat" sensitive information \cite{pew-2014-privacy_post_snowden}.  One reason marketers may argue that online data collection is acceptable is that consumers are making a tradeoff between privacy and discounts; yet a 2015 study found that 91\% of respondents \emph{disagreed} with the statement ``If companies give me a discount, it is a fair exchange for them to collect information about me without my knowing it" \cite{turow-2015-tradeoff_fallacy}.  The general trend established by these surveys is not just that most users value privacy, but that the current state of privacy online represents an area of significant anxiety.  Indeed, research into the mechanisms of tracking highlight that there is serious cause for concern.

	\subsection{Mechanisms}
	Computer security research has explored many novel methods of compromising user privacy.  This has included detecting the if a user had visited a page based on the color of a given hyperlink \cite{jackson-2006-protecting}, analyzing Flash ``cookies" which survive user deletion \cite{ayenson-2011-flash}, the increasingly popular fingerprinting techniques discussed above \cite{eckersley-2010-unique,nikiforakis-2013-cookieless,acar-2013-fpdetective}, as well as other uses of Javascript \cite{jang-2010-empirical}.  The common thread among such research is to focus on ways in which a browser may be used to extract more information about the user than is typically provided by HTTP request headers.  This focus on what is called the ``client-side" (i.e. the user's computer rather than the server) is born of the fact that researchers have long assumed ``that sites are not able to reliably track users using just their IP address and user-agent string" \cite{jackson-2006-protecting}.  In other words, this means that the basic information detailed above from an HTTP request is insufficient to track users with the same level of accuracy as other methods such as cookies or fingerprinting.  However, this assumption has been proven wrong.

	Even without direct browser manipulation, Yen et al. have demonstrated that solely by analyzing HTTP headers on a multi-million point dataset they were able to reach a user identification rate of 80\%, which is ``similar to that obtained with cookies" \cite{yen-2012-host}.  This finding is significant because it demonstrates that third-party HTTP requests \emph{by themselves} hold similar potential for compromising user privacy as the more commonly studied client-side techniques.  A single HTTP request is unlikely to result in user identification, but thousands of such requests may be correlated to a given user.  Such techniques happen on the servers of corporations, and are largely hidden from analysts of client-side code.

	\subsection{Measurement}
	The third general area of research focus has been in measurement.  In this case, the interest of the researcher is in selecting a population of websites and essentially conducting a census of tracking mechanisms, third-party requests, and data flows.  Much of the pioneering work in this area has been conducted by Krishnamurthy and Wills.  In 2006 they advanced the concept of the ``privacy footprint" which they describe as a means to analyze the ``diffusion of information about a user's actions by measuring the number of associations between visible nodes via one or more common hidden nodes" \cite{krishnamurthy-2006-generating}.  What Krishnamurthy and Wills describe as ``hidden nodes", may be more widely included under the umbrella of third-party requests.  This work was followed up several years later with a longitudinal study \cite{krishnamurthy-2009-privacy}.  The data presented in this article updates and advances Krishnamurthy and Wills findings with a larger sample.
	
	While Krishnamurthy and Wills have done an excellent job in the area of measurement, they are far from the only researchers exploring the topic.  Castelluccia and colleagues have recently analyzed ``the flows of personal data at a global level" \cite{Castellucia-2013-dataharvesting2}, and found intriguing differences in the nature, scope, and ownership of tracking mechanisms around the world.  Other researchers have developed desktop browser add-ons \cite{roesner-2012-detecting,mayer-2012-third} which they have used to detect the presence of a variety of tracking mechanisms.  Some of the most recent measurement literature focuses on fingerprinting specifically \cite{acar-2013-fpdetective}.  Additional research has revealed extensive tracking on websites containing health information \cite{libert-2015-healthtracking}.  A common theme among all measurement research is that the amount of tracking on the web is increasing, and shows no signs of abating.

\section{Research Questions}
	This article seeks to update and advance knowledge in three main areas: the general size and nature of the hidden web, the corporate ownership patterns which affect user privacy, and the degree to which state surveillance is facilitated by commercial tracking mechanisms.  In regards to general questions on the size and scope of the web tracking ecosystem, several issue are explored in depth, among them are the number of pages which leak user data, how many entities users are exposed to in tandem, and how this varies among different countries.  Furthermore, third-party elements are investigated in order to determine if they are visible to the user, or represent potential attempts at covert tracking.  Second, and of central interest to this study, the corporations who are responsible for tracking users online are discovered.  This focus sheds light on the larger social and political-economic considerations involved in web tracking.  Finally, internal documents released by former U.S. National Security Agency (NSA) contractor Edward Snowden, have revealed that the spy agency has leveraged commercial tracking tools in order to monitor users.  Thus, this research project also seeks to detect commercial tracking mechanisms known to be co-opted by the NSA.

\section{Methodology}
	Following the findings of Yen et al. described above, it is clear that third-party HTTP requests provide an excellent unit of analysis for examining the extent of tracking on the web.  Such requests are potentially as revealing and invasive as cookies, and are a basic prerequisite of every single family of tracking techniques.  Regardless of what cookies are set, or what ``fingerprints" are taken from a browser, a connection between the user and the server of the tracker needs to be established via HTTP.  By examining HTTP requests in bulk, a picture of data flows emerges.  On a sufficiently large data set those requests which have a benign function fade into the long tail of data, and those which happen with great regularity are easily spotted.  Therefore, this research faces two basic problems: what sites form a sufficiently large population, and how to analyze third-party HTTP requests on each site.  
	
	The best population of sites was determined to be the top one million sites list published by Alexa\footnote{http://s3.amazonaws.com/alexa-static/top-1m.csv.zip}\footnote{https://support.alexa.com/hc/en-us/articles/200449834-Does-Alexa-have-a-list-of-its-top-ranked-websites-}. Alexa is a subsidiary of Amazon who provides website traffic metrics and rankings derived from a ``panel of toolbar users which is a sample of all [I]nternet users"\footnote{https://alexa.zendesk.com/hc/en-us/articles/200449744-How-are-Alexa-s-traffic-rankings-determined-}.  The degree to which Alexa's data accurately represents popularity on the web is debatable as Alexa Toolbar users are by no means a random sample of all Internet users.  However, it has become common practice for researchers to use the Alexa list in the absence of other, better lists \cite{krishnamurthy-2006-generating,krishnamurthy-2009-privacy, jang-2010-empirical, mayer-2012-third, roesner-2012-detecting, Castellucia-2013-dataharvesting2, nikiforakis-2013-cookieless, acar-2013-fpdetective}.  For this study the full Alexa list was downloaded in May of 2014.

	In order to detect tracking on the sites selected, the webXray software (version 1.0) was used.  webXray is written primarily in Python and works in several steps.  First, a list of website addresses is given to the program.  This list is processed to ensure all of the addresses are properly formatted, are not links to common binary files (such .pdf or .xls), and are not duplicates.  Next, each site is relayed via the Python \emph{subprocess} module to a command-line instantiation of the headless web browser PhantomJS.  In this context, ``headless" refers to the fact that the browser runs on a command line and does not require a graphical user interface; this makes PhantomJS ideal for deployment on a ``cloud"-based virtual machine.  PhantomJS takes as arguments a web address and a Javascript program which is responsible for loading the website, processing the page title and metadata, collecting cookies, and detecting both HTTP requests and HTTP received events.  In order to successfully follow multiple redirects and allow elements to download, PhantomJS is given 30 seconds to complete loading the page.  The results are passed back to the Python script as JSON data and processing resumes.  Data about the page, such as title and meta description, are stored in a database, and then HTTP requests are examined.

	webXray utilizes the Mozilla Public Suffix List \footnote{http://publicsuffix.org/} in order to determine the domain of each HTTP request (sub-domains are ignored).  Requests which have the same domain as the page address are ignored (e.g. ``http://example.com" and ``http://images.example.com/header.png"), whereas requests with different domains are entered into a database of third-party elements (e.g. ``http://example.com" and ``http: //www. google-analytics.com/ \_\_utm.gif").  This process is repeated for storing third-party cookies.  Request strings are additionally parsed to remove argument data from the domain and element.  For example, the following string:
	\begin{verbatim}
	http://sub.example.com/tracking_pixel.png?id=8675309
	\end{verbatim}
	
	 is broken up as having the domain ``example.com", the element ``tracking\_pixel.png", the arguments ``?id=8675309", and the file extension ``png".  File extensions of elements (e.g. ``png", ``js", ``css") are used to catalog elements by type (e.g. ``image", ``Javascript", ``Cascading Style Sheet").
	
	Finally, after a scan is completed the top 100 domains are identified and ownership information is manually determined.  The process primarily relies on using the \emph{whois} utility to look for domain registration records.  In cases where whois records are unhelpful the domain or element is loaded into a browser in order to follow redirects to a parent site.  If this fails, web searches are used to find ownership.  When smaller companies are found, the website Crunchbase is consulted to determine if they are subsidiaries or acquisitions of larger companies.  Great care in this process is taken to ensure ownership findings are accurate and the ``About'' pages of all companies are consulted as part of the process.  One additional barrier is the fact that many companies have several domains.  For example, Google has by far the most domains, with several hundred unique entries listed on Wikipedia\footnote{https://en.wikipedia.org/wiki/List\_of\_Google\_domains}; the names of these domains range from the transparent (``google.com", ``google.fr") to the opaque (``1e100.net", ``ggpht.com").  Adobe was found to be using many domains as well - again, some obvious (``adobe.com"), and some not (``2o7.net").  This pattern of household brand-names using unfamiliar domains held fast for a number of companies including Twitter, Facebook, and Amazon.  It should be noted that there is nothing particularly nefarious about companies using unfamiliar domains to host content, only that it presents a barrier during the analysis process.  In total, over 500 distinct domains were traced back to 140 different companies.  In order to reduce false-positives, domains which are requested by only a single site are ignored in the final analysis.  For example, if the site ``http://example.com" was the \emph{only} site to initiate a request to the domain ``example-images.com", it is ignored.  This approach leaves only domains which can conceivably link visitors between two or more sites in the analysis pool.

	\begin{table*}[h]\footnotesize
		\caption{Findings Summary}
		\begin{center} 
		\begin{tabular}{| l | l | l | l | l | l | l |}
			\hline
			Rank	&	TLD	&	N	&	\% W/3PE	&	Ave. Domains Contacted	&	\% W/Cookie	&	\% W/JS\\
			\hline
			- & *	&	950,489	&	86.9	&	9.47	&	62.9	&	82.99\\
			\hline
			1	&	com	&	517,085	&	88.47	&	10.2	&	64.66	&	84.5\\
			\hline
			2	&	net	&	49,471	&	83.03	&	10.2	&	61.58	&	77.65\\
			\hline
			3	&	org	&	37,568	&	85.73	&	7.81	&	58.23	&	81.28\\
			\hline
			4	&	ru	&	37,560	&	94.33	&	8.48	&	89.42	&	88.72\\
			\hline
			5	&	de	&	32,447	&	85.34	&	7.52	&	52.87	&	79.26\\
			\hline
			6	&	uk	&	17,976	&	87.47	&	9.11	&	61.09	&	85.02\\
			\hline
			7	&	br	&	15,420	&	87.85	&	11	&	62.33	&	85.65\\
			\hline
			8	&	jp	&	14,478	&	82.05	&	6.87	&	48.08	&	79.52\\
			\hline
			9	&	pl	&	12,627	&	90.04	&	7.75	&	57.87	&	87.96\\
			\hline
			10	&	in	&	12,271	&	81.79	&	10.35	&	57.86	&	77.14\\
			\hline
			.. & & & & & & \\
			\hline
			43 & edu	&	2,416	&	89.74	&	5.67	&	47.64	&	88.49\\
			\hline
			.. & & & & & & \\
			\hline
			65 & gov	&	762	&	86.48	&	3.61	&	39.63	&	84.65\\
			\hline
		\end{tabular}
		\end{center}
	\end{table*}

\section{Findings}

	\subsection{General Trends}
	The home pages of the most popular 990,022 websites as measured by Alexa were analyzed in May, 2014.  96\% of sites (950,489) were successfully analyzed and 4\% failed to load.  This failure may be attributed to the top sites lists containing sites which were not available at the time of analysis (other webXray tests yielded success rates of over 99\%, indicating the software is not at fault).  The pages analyzed yielded a set of 7,564,492 unique cookies, 21,214,652 unique element requests, and encompassed 1,056,533 distinct domains.  The number of pages which initiated requests to third-parties was 832,349, representing 88\% of the total.  Further analysis determined that sites which \emph{do} make requests to third parties contact 9.47 distinct domains on average - indicating that not only is user data being frequently shared, but it is often being shared with many parties simultaneously.  62.9\% of sites spawn third-party cookies which may be used to track users using traditional methods, and 82.99\% of sites include third-party Javascript which may represent the increasing trend towards fingerprinting and other advanced techniques.  It should be noted that since this study used a specified population rather than a statistical sample, the confidence for these numbers is 100\%, but these results \emph{only} represent the Alexa top one million - \emph{not} the entire web.
		
	In addition to the top level trends, sub-analyses were conducted on the 10 most common top-level domains (TLDs) as well as U.S. educational (edu) and government (gov) sites.  The top TLDs include the well-known com, net, and org as well as several country-code top-level domains (ccTLDs) such as Russia (ru), Germany (de), United Kingdom (uk), Brazil (br), Japan (jp), Poland (pl), and India (in).  Sites ending in a ccTLD have been registered within a given country - however, popular sites within a given country often include com and org sites, so this is \emph{not} a perfect proxy for what a user from a given country is exposed to (i.e. do not assume Russians only visit \emph{ru} sites).  When comparing data among TLDs, there was not much range in the percentage of sites which had a third-party element: Russian domains had the most with 94.33\%, and Indian domains had the least with 81.79\%.  However, in regards to the average number of domains contacted, there was a large spread with Brazilian sites contacting 11 on average and U.S. government sites contacting 3.61.  There was also a wide disparity in the percentage of sites spawning third-party cookies: Russia was again in first place with 89.42\% of sites spawning cookies, whereas only half as many U.S. government sites (39.63\%) did so.  Finally, there was not much difference in the percentage of sites with third-party Javascript: at the high end were Russian sites with 88.72\%, and at the low end were Indian sites with 77.14\%.  These findings are presented in greater detail in Table 1.

	\subsection{Types of Requested Elements}

	In order to determine current trends in tracking mechanisms, the top 100 requested elements were categorized according to six major types based on file extensions.  Javascript is by far the most prevalent, being requested in 36\% of cases.  Javascript is a popular development language, and many uses of Javascript are wholly benign, yet still generate HTTP request records and may potentially be used for advanced fingerprinting techniques. The analysis of Javascript code is outside the scope of this article, and this prevalence is noted in order to highlight trends, not to suggest all Javascript is designed for fingerprinting. Nevertheless, Javascript elements may pose a greater potential privacy risk than inert elements such as images.  Pages with the ability to serve dynamic content, such as PHP and CGI scripts, make up 9\% of the total.  Images, such as tracking pixels, have a history stretching back to the early days of the web and are still popular, comprising 7\% of the total.  Fonts and Cascading Styles Sheets (CSS) together make up 5\% of requests.    Two requests for structured data (JSON) were found in the top 100.  38\% of requested elements were not categorizable for they lacked recognizable file name extensions.

	\subsubsection{Images}	
	Of the seven identified element types above, six are in no way visible to the user and will not provide any clue that she or he is being tracked.  However, images could presumably offer a means for a user to determine there are additional parties present.  Yet, as Table 2 details, only four of the top ten images are visible, and only two of these have any indication of what company owns them.  The lack of visibility or ownership attribution highlights that these images are designed to avoid detection by users, and are  likely a tracking tool.  However, it is well worth noting that even if a given image is invisible, it may be accompanied by text on the screen which tells the user that tracking is occurring.  The most likely case for this happening is if the page is also setting a cookie and the user is in an European country covered by the “cookie” law (which globally, is not he majority of users).  Based on the methodology used in this study it is not possible to account for such cases should they exist.  That said, half of the top ten images belong to Google, including the most requested image, the Google Analytics tracking pixel.  This image is found on 46.02\% of sites, is only 1x1 pixels large, and is utilized solely for tracking purposes.  The second most requested image is the Facebook ``Like Button".  This image is both visible and clearly indicates the owner (albeit via iconography rather than text); it's prevalence, on 21.34\% of sites, demonstrates that the visual language of Facebook is having a significant influence on the Alexa top sites.  It should be noted that while this is a Facebook image, the domain hosting it is ``akamaihd.net".  This domain belongs to the Akamai company which provides hosting services for major brands around the world.

	\begin{table*}[h]\footnotesize
		\caption{Top Image Characteristics}
		\begin{center} 
		\begin{tabular}{| l | l | l | l | l | l | l |}
			\hline
			Rank & \% Sites & File Name & Domain & Company & Visible & Indicates Owner \\
			\hline
			1 & 46.02\% & \_\_utm.gif & google-analytics.com & Google & N & N \\
			\hline
			2 & 21.34\% & LVx-xkvaJ0b.png & akamaihd.net & Akamai/Facebook & Y & Y\\
			\hline
			3 & 6.97\% & nessie\_icon\_tiamat\_white.png & googlesyndication.net & Google & Y & N\\
			\hline
			4 & 5.16\% & \_\_utm.gif & doubleclick.net & Google & N & N\\
			\hline
			5 & 3.81\% & g.gif & wordpress.com & Wordpress & N & N\\
			\hline
			6 & 3.45\% & vendor.gif & reson8.com & Resonate & N & N\\
			\hline
			7 & 2.57\% & icon18\_wrench\_allbkg.png & blogblog.com & Google & Y & N\\
			\hline
			8 & 1.78\% & demconf.jpg & demdex.net & Adobe & N & N\\
			\hline
			9 & 1.73\% & small-logo.png & google.com & Google & Y & Y\\
			\hline
			10 & 1.69\% & pixel.gif & amazonaws.com & Amazon & N & N\\
			\hline
		\end{tabular}
		\end{center}
	\end{table*}

	\subsection{Corporate Ownership}
	The major contribution of this article is to reveal the corporate ownership of tracking mechanisms.  This focus is vital as the very reason that most tracking occurs is for economic motives.  In order to understand the phenomena of web tracking, and to craft meaningful policies regulating it, it is necessary to know the actual extent and reach of the corporations which may be subject to increased regulation.  The most striking finding of this study is that 78.07\% of websites in the Alexa top million initiate third-party HTTP requests to a Google-owned domain.  While the competitiveness of Google is well known in search, mobile phones, and display advertising, its reach in the web tracking arena is unparalleled.  The next company, Facebook, is found on a still significant 32.42\% of sites, followed by Akamai (which hosts Facebook and other companies' content) on 23.31\% of sites, Twitter with 17.89\%, comScore with 11.98\%, Amazon with 11.72\%, and AppNexus with 11.7\%.  As Figure 1 demonstrates, these seven companies dwarf the impact of the next 43 companies.  The distribution of companies by percentage of sites tracked has a marked peak, which quickly descends into a long tail of 33 companies which track between 1-4\% of sites.

	\subsection{Commercial Tracking and State Surveillance}
	
	In December, 2013, \emph{The Washington Post} reported that ``The National Security Agency is secretly piggybacking on the tools that enable Internet advertisers to track consumers, using `cookies' and location data to pinpoint targets for government hacking and to bolster surveillance" \cite{soltani-2013-nsa_google}.  More specifically, internal NSA documents leaked to the \emph{Post} by former NSA contractor Edward Snowden revealed that a Google cookie named ``PREF" was being used to track targets online.  Additional documents provided to \emph{The Guardian} by Snowden detailed that another Google cookie (DoubleClick's ``id"), was also used by the NSA; in this case to attempt to compromise the privacy of those using anonymity-focused Tor network \cite{guardian-2013-tor_stinks}.  
	
	As noted above, in the course of analyzing the Alexa top one million websites, 7,564,492 unique cookies were collected.  Investigation of this cookie collection revealed 81,699 pages (8\% of the total) spawning cookies with the name ``PREF" set by Google-owned domains, and an additional 180,212 pages (19\% of the total) spawning cookies with the name ``id" from the (also Google-owned) DoubleClick domain.  Given the fact that the Snowden disclosures have often raised more questions than answers, it is unclear if the identified cookies are being used to surveil users today, and companies like Google are working admirably hard to improve security.  However, it is clear that the widely deployed tracking mechanisms identified in this article are of interest to more than just advertisers.  In the absence of meaningful oversight, such mechanisms are ripe for abuse - not only by the United States intelligence agencies, but by numerous states around the globe.

	\subsection{Limitations}
	
	While the methodology described above performs very well at a large scale, it is not without its flaws, and the counts presented in this article may fall short of the actual number of third-party HTTP requests being made.  These findings therefore represent a \emph{lower bound} of the amount of third-party HTTP requests, with the actual amount likely being slightly higher.  However, given the extensive prevalence of third-party HTTP requests, this constraint only serves to highlight the magnitude and scope of the findings.

	Potential under-counting may be mainly attributed to the nature of the investigation itself: third-parties have no desire to be detected by users in the vast majority of cases.  As previous research into mechanisms has demonstrated, advertisers go to extreme lengths to devise exotic techniques designed to circumvent user preferences in order to track their activities.  For example, Google recently paid out \$39.5 million in federal and state settlements over its use of techniques which circumvented cookie preferences on the Safari web browser \cite{fung-2013-googlesettlement}.  While nobody went to jail, and Google admitted no fault, it is becoming clear that advertisers regularly skirt the border between violating user preferences and outright violating the law.  For this reason, any census of third-party requests on the web is bound to be subject to an under-count as those profiting from web tracking prefer to remain in the shadows.
	
	An additional weakness is that PhantomJS does not support Adobe Flash, Microsoft Silverlight, or other browser-based plugins.  Requests to download such content are not detected, and if such content subsequently initiates additional requests those are not recorded either.  Historically, Flash has been a popular means of tracking on the web.  However, Flash is not supported by the vast majority of smartphones and tablets.  Given the fact that industry estimates of mobile browser usage now stand at 37.12\% globally\footnote{http://gs.statcounter.com/\#all-comparison-ww-monthly-201407-201507}, not only are Flash elements failing to reach over a third of users, it is apparent that new developments in tracking will likely not rely on Flash.  It may be too early to declare the end of Flash-based tracking, but the conclusions of this study are not significantly skewed by this methodological weakness.

	A final reason there may be an undercount is the rapid rate of ingestion.  Given that webXray downloads over 10,000 pages per hour, the volume of requests from the IP address of the server running webXray will stand out as highly irregular.  Many web masters seek to limit automated scanning behavior by blocking overly-active IP addresses.  Therefore, it is possible that some third-party domains would have automatically blocked the IP address, yielding an under-count.  This is one of many trade-offs made to achieve high speed and low cost, and the above limitations are detailed in order to account for the most obvious weaknesses in the technique.  
	
	A final note is that the analysis was conducted in the United States where there are no laws which dictate websites must notify users of cookies, as in the European Union.  Thus, elements which are wholly invisible in the United States and elsewhere may come with visual indications for EU users.  No research methodology is perfect, and the enumerated problems represent the type of relative uncertainty which is common in all research.  None of the problems with webXray serve to significantly undermine the extensive findings detailed above.

\section{Discussion: Do Not Track}
	
	Of the top 100 third-party elements discovered in this study, only seven were images; of these images, only two provided any indication of who the images were owned by.  This means that 98\% of the identified tracking elements may never alert users to their presence.  Furthermore, the 2\% which indicate ownership do not advertise their purpose - only their presence.  Thus, the question we must face is that if attitudes research has proven users do not want to be tracked, and such tracking in wide-spread and largely undetectable, how can we bring user expectations and corporate behavior inline?  How can we eliminate the socially ``intolerable" from the present ad-supported web?  There are three main solutions which are being pursued at present: industry ``opt-out'' mechanisms, browser add-ons, and the ``Do Not Track" (DNT) mechanism.  The first two solutions are ineffectual to varying degrees, and the third will need the power of legislation in order to be respected by corporations.
	
	While it is close to impossible for non-EU users to see they are being tracked, they have the possibility to infer as much from the presence of advertisements in their browser.  Many such advertisements come with a very small blue arrow which when clicked will eventually bring the user to the ``online home of the Digital Advertising Alliance's (DAA) Self-Regulatory Program for Online Behavioral Advertising".\footnote{http://www.aboutads.info}  On this websites a user may opt-out of targeted advertising.  However, while most reasonable users would assume opting-out means they \emph{will not be tracked}, the DAA interprets it to mean that users ``opt out from receiving interest-based advertising".\footnote{http://www.aboutads.info/choices/}  This means that users may \emph{still be tracked}, they just will not see tailored advertisements.  This practice is misleading at best and deceptive at worst.  Furthermore, the opt-out mechanism requires the user to set a large array of cookies on her or his browser.  Thus, if a user has been protecting their privacy by turning off third-party cookies, she or he must \emph{turn them back on} and potentially be exposed to \emph{more} tracking as not all trackers are members of the DAA.  The industry solution is insufficient at best and intentionally misleading at worst.

	The next means of protection, the use of browser add-ons, is vastly more effective at preventing tracking, but comes with several problems.  First, the only means a user has for knowing that such mechanisms are available is by reading news stories which highlight the add-ons, or by browsing lists of popular add-ons for various browsers (privacy add-ons routinely make top-ten lists).  A user must therefore actively seek out this complex remedy and spend a significant amount of time learning about how their browser works.  For some this may be trivial, but for many it is a difficult proposition.  Even assuming a user is able to install an add-on, they are offered incomplete protection: popular add-ons such as Ghostery and Ad-Block+ rely on blacklists of known tracking elements and fail to prevent all tracking.  A Firefox add-on, RequestPolicy \cite{samuel-2009-requestpolicy}, does block all third-party requests, but at the expense of breaking the rendering mechanisms of many pages, leaving the average user with a confused mess of text on her or his screen.  While these add-ons do indeed offer a significant degree of protection, they also place the burden on users and highlight the blame-the-victim mentality at the core of the issue.  Furthermore, these add-ons are not available on the stock browsers included on Android phones and tablets, though recent additions to Apple iOS 9 have introduced content blocking.
	Given the shortcomings of the above two solutions, it would seem that the simplest way to reduce unwanted tracking would not be setting hundreds of cookies, or installing cumbersome browser add-ons, but by having a clear way for users to signal to corporations that they should not be tracked.  But what is the best way to send this signal?  Once again, the lowly HTTP request is the answer.  It is now possible to configure all major web browsers (Internet Explorer, Safari, Chrome, Firefox, and Opera) to send a message within the body of a request which indicates the user does not want to be tracked; when the aforementioned ``user-agent" is sent to a server, another field, ``DNT" may be sent as well.  DNT may be set to ``1", indicating a user does not want to be tracked, or ``0", indicating they would like to be tracked; if there is no value set, no conclusions about a user's preference may be drawn \cite{mayer-2011-DNT}.  It would seem the problem is therefore solved: a user may now easily tell trackers to leave them alone by changing a single browser setting.
	
	However, DNT comes with no enforcement mechanism - it is merely a polite request, and trackers may ignore it as they choose - and indeed they do.  For example, comScore, the fifth most prevalent tracker, states in their privacy policy that ``comScore.com does not react to [i.e. ignores] Do Not Track signals".\footnote{http://www.comscore.com/About-comScore/Privacy-Policy}  Investigation of the policies of the top ten corporations with tracking elements shows that nine of them do not respect the DNT header.  One company, Yahoo!, respected DNT for a time, but on April 30th, 2014 reversed course and stated that ``As of today, web browser Do Not Track settings will no longer be enabled on Yahoo".\footnote{http://yahoopolicy.tumblr.com/post/84363620568/yahoos-default-a-personalized-experience}  The \emph{only} company in the top ten to respect the DNT header is Twitter.  According to Twitter, ``When you turn on DNT in your browser, we stop collecting the information that allows us to tailor suggestions based on your recent visits to websites that have integrated our buttons or widgets"\footnote{https://support.twitter.com/articles/20169453-twitter-supports-do-not-track}.  It is important to note that Twitter follows the reasonable definition of opt-out, namely user data is not \emph{collected}, as opposed to the DAA's solution of tailored advertising not being \emph{shown}.  Twitter's position is admirable and represents the standard to which other companies should aspire.

	Given that 90\% of the top ten tracking companies ignore DNT, it seems that the standard is a failure.  However, this does not need to be the case.  As mentioned earlier, Hoofnagle et al. have determined that 60\% of those surveyed supported the DNT standard.  If these users were able to have their wishes given the force of law - as should be possible in democracies - the behavior of Twitter would not represent a benevolent exception, but a legally enforced rule.  The fact of the matter is that the technical groundwork exists, public desire exists, and all that is still needed is for law makers to fulfill their duty to side with citizens against unwelcome corporate practices.
		
\section{Conclusion}

	Three things have hopefully been established by this article: first, Westin's socially ``intolerable" use of technology has come to pass; second, Senator Church's warning of an ``abyss" of technologically advanced state surveillance was prophetic; and third, there \emph{is} a way out.  By using a quantitative approach towards the issue of web tracking, it has been determined that user privacy is widely compromised by numerous parties in tandem, users often have limited means to detect these privacy breeches, a handful of powerful corporations receive the vast majority of user data, and that state intelligence agencies such as the NSA may be leveraging this commercial surveillance infrastructure for their own purposes.  Furthermore, current privacy protections are wholly inadequate in light of the scale and scope of the problem.  Despite this, by having singled out the corporations who covertly observe users, it is now possible to identify the degree to which these companies abide by expressed user preference, or fail to do so.  A full exploration of specific policy implementations is beyond the purview of this research, but the measures contained herein may be used by the policy community to inform, support, and advance solutions which ensure that DNT moves from a polite request to an enforceable command.


\bibliographystyle{abbrv}
\bibliography{/Users/tlib/Continuity/PDF/mybigbib}


	\begin{landscape}
		\vspace*{\fill}
		\begin{figure}[H]
			\includegraphics[scale=.55]{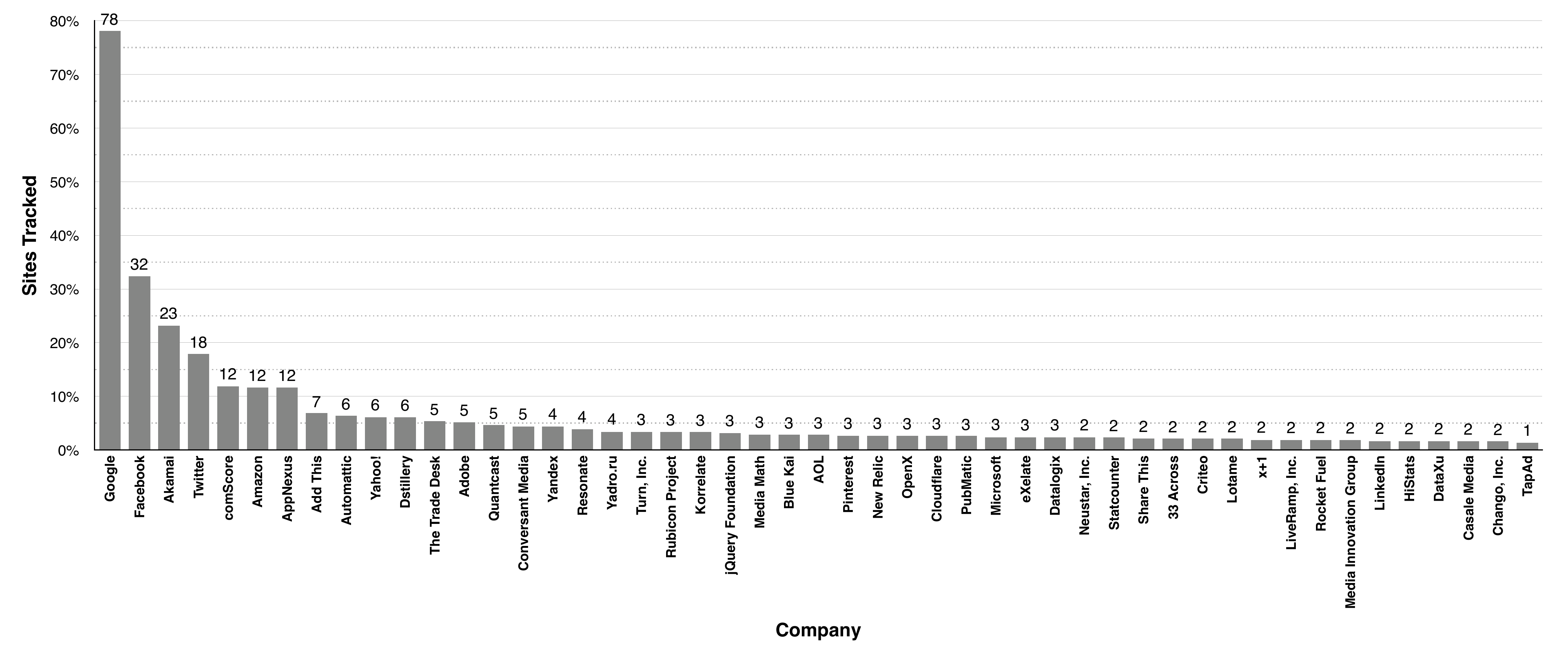}
			\caption{\emph{Percentage of Sites Tracked by Top 50 Corporations}}
		\end{figure}
	\end{landscape}

\end{document}